# COHERENT QUASIPARTICLE TRANSPORT IN GRAIN BOUNDARY JUNCTIONS EMPLOYING HIGH-T$_C$ SUPERCONDUCTORS

*F. Tafuri,[1,2] A. Tagliacozzo,[2] D. Born,[2] D. Stornaiuolo,[2] E. Gambale,[1]*
*D. Dalena,[1] P. Lucignano,[2] B. Jouault,[3] F. Lombardi,[4] A. Barone,[1] and B.L. Altshuler [5]*

[1]Dip. Ingegneria dell'Informazione, Seconda Università di Napoli, 81031 Aversa (CE), Italy
[2]Dip. Scienze Fisiche, Università di Napoli Federico II, 80125 Napoli, Italy
[3]GES, UMR 5650, Université Montpellier II 34095 Montpellier Cedex 5, France
[4]Dept. Microelectronics and Nanoscience, MINA, Chalmers University of Technology, 41296 Göteborg, Sweden
[5]Physics Dept. Columbia University, New York NY 10027; NEC Laboratories America INC, 4 Independence Day, Princeton, NJ 08554

**ABSTRACT**

Magneto-fluctuations of the normal resistance $R_N$ have been reproducibly observed in YBa$_2$Cu$_3$O$_{7-d}$ biepitaxial grain boundary junctions at low temperatures. We attribute them to mesoscopic transport in narrow channels across the grain boundary line, occurring even in the presence of large voltage drops. The Thouless energy appears to be the relevant energy scale. Possible implications on the understanding of coherent transport of quasiparticles in HTS and of the dissipation mechanisms are discussed.

**1. INTRODUCTION**

The progress in the realization of high quality high critical temperature superconductors (HTS) junctions is steady and regards both technological issues and the study of the fundamental transport mechanisms [1-3]. Grain boundaries still represent privileged systems to study the physics of the Josephson effect in HTS structures [2,3]. Relevant insights can be obtained by investigating quantum and macroscopic quantum effects in appropriate systems [3-5]. Recently, high quality YBa2Cu3O7-d (YBCO) grain boundary (GB) Josephson junctions (JJs) have given the first evidence of macroscopic quantum behaviors, coherence and dissipation [4-5].

We report on transport measurements of YBCO biepitaxial (BP) GB junctions (see scheme in Fig. 1), which give evidence of conductance fluctuations (CFs) in the magneto-conductance response. These results prove the appearance of mesoscopic effects in the unusual energy regime $k_B T \ll e_c < eV < D$. Here $k_B T$ is the thermal energy at temperature $T$, $e_c$ is the Thouless energy (see below), $V$ is the applied voltage and $D$ is the superconducting gap (~ 20 meV for YBCO [1]). These effects have to be ascribed to the presence of intrinsic narrow channels appearing in the formation of the grain boundary. Transmission Electron Microscope analyses have confirmed the existence of about 50 nm wide microbridges in these junctions. The existence of narrow channels (filamentary structure) embedded in a basically insulating matrix has been claimed to be a common feature of GB junctions, with relevant influence on their transport properties [6]. Nevertheless, definitive answers on their actual role are still missing [2,3]. Transport probably results from the interplay of various processes, strongly dependent on the barrier microstructure, which acts as filter for the various processes.

The analysis presented in this manuscript gives additional insights on the understanding of the relaxation processes of low energy quasiparticles in HTS systems, as we will discuss below. Through a "direct" measurement of the Thouless energy, the energy scale we found sets a lower bound to the relaxation time at low temperatures of the order of picoseconds. This conclusion is consistent with the successful observation of macroscopic quantum effects in GB HTS Josephson junctions. It also supports proposals of quantum devices based on HTS junctions, encouraging novel approaches which exploit the robust mesoscopic features of nanochannels across GBs.

At low temperatures, quantum coherence can be monitored in the conductance G of a normal metallic sample of length $L_x$ attached to two reservoirs [7-10]. The electron wave packets that carry current in a diffusive wire have minimum size of the order of $L_T > L_x \gg l$. Here $l$ is the electron mean free path in the wire and $L_T$ is the thermal diffusion length $\sqrt{\hbar D / k_B T}$ ($D$ is the diffusion constant).

The first inequality is satisfied at relatively low temperatures as far as $k_B T \ll e_c \sim hD/L_x^2$. At low voltages ($eV \ll e_c$), the system is in the regime of universal conductance fluctuations: the variance $\langle \delta g^2 \rangle$ of the dimensionless conductance $g = G/(2e^2/h)$ is of order of unity.

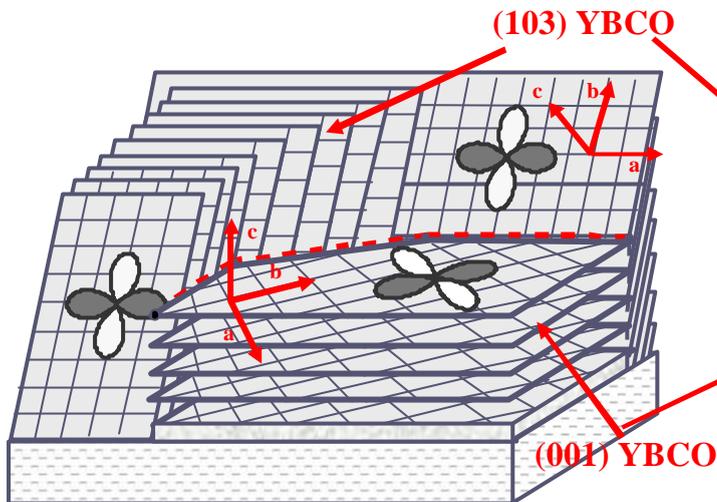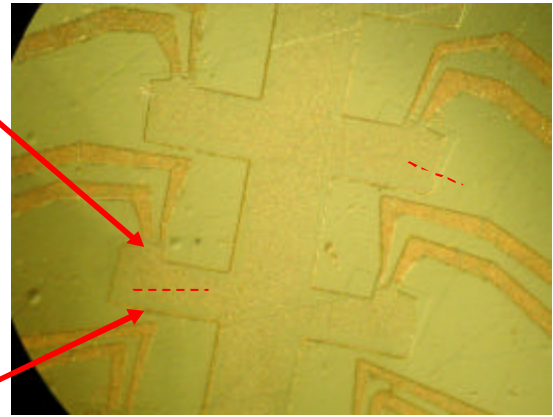

Fig. 1.a) Sketch of the biepitaxial grain boundary structure for three different interface orientations. We have indicated the two limiting cases of 0° and 90°, and an intermediate situation defined by a generic angle θ.

Fig. 1b) Optical image of various grain boundary junctions with different interface orientations. The junctions are 3 and 6 μm wide respectively. Two examples of grain boundary lines are marked by a red dashed line.

## 2. EXPERIMENTAL TECHNIQUES AND RESULTS

The BP JJs are obtained at the interface between a (103) YBCO film grown on a (110) SrTiO$_3$ substrate and a c-axis film deposited on a (110) CeO$_2$ seed layer (Fig. 1). The presence of the CeO$_2$ produces an additional 45° in-plane rotation of the YBCO axes with respect to the in-plane directions of the substrate. By suitable patterning of the CeO$_2$ seed layer, the interface orientation can be varied and tuned to some appropriate transport regime (in Fig. 1 we indicated the two limiting cases of 0° and 90°, and an intermediate situation defined by the angle $q$, in our case $q = 60°$) [4,5,11,12]. We have selected junctions with sub-micron channels, as eventually confirmed by Scanning Electron Microscope analyses, and measured their *I-V* curves, as a function of the magnetic field *H* and temperature *T*.

The transport measurements, with a sample temperature down to 0.3 K (3He evaporation cryostat), are performed using a standard four terminal setup. In case of the current-voltage (IV) characteristics the current is driven by a voltage source in series with a 10 kΩ resistance. The voltage is recorded by an oscilloscope via a preamplifier on top of the cryostat. In case of differential resistance measurements, the ac-current (7.8Hz) is driven by a ac-voltage source in series with a 10kΩ resistance. The differential voltage is measured by lock-in also via the preamplifier. The electrical lines to the sample are equipped with copper powder microwave filters (low pass, cut-off frequency about 10 GHz) and discrete RC-filters (low pass, 1.6 MHz). They are both placed at the 1.5 K temperature stage of the cryostat. The magnetic screening is guaranteed by two Cryoperm screens and a two layer (Pb/Nb) superconducting screen inside the liquid He dewar.

We have investigated various samples, but here we focus mostly on the junction characterized by the Fraunhofer magnetic pattern shown in Fig. 2, where I-V curves are reported as a function of the magnetic field. In this case, it is more likely that only one uniform superconducting path is active. The critical current $I_c$ oscillations point to a flux periodicity which is roughly consistent with the typical size (10-100 nm) that we expect for our microbridge from structural investigations.

In our case, a linear branch typical of the Resistively Shunted Junction behavior starts at $V = 5$ mV. The normal state resistance $R_N$ can be reliably estimated between 10 mV and 15 mV. The average normal state resistance $R_N$ over the whole magnetic field range is ~ 182 Ω. We choose $V \sim 12$ mV for deriving the $R_N$ vs. *H* curve, that is reported in Fig. 3 at the temperature $T = 300$ mK in the interval [-100 G, 100 G]. Conductance fluctuations, as those shown in Fig. 3, become appreciable at temperatures below 900 mK, in the whole magnetic field range. Their amplitude is $\sqrt{<\Delta G^2>} \approx 0.07 \overline{G_N}$ (with $\overline{G_N} \equiv \overline{R_N^{-1}}$) at 300 mK and are more than one order of magnitude larger than the noise. The fluctuations are nonperiodic, and for sure not related to the $I_c(H)$ periodicity, as shown in Fig. 3, and have all the typical characteristics of mesoscopic fluctuations. Above 1.2 K thermal fluctuations dominate. In Fig. 3, as a reference, we have also plotted the critical current vs magnetic field H. The different periodicity of $I_c$ and $R_N$ vs H have been verified on a larger H scale up to 600 Gauss, and will be discussed elsewhere. We recall that the possibility to trap magnetic flux at higher fields for Josephson junctions determines some constraints for the maximum applied field.

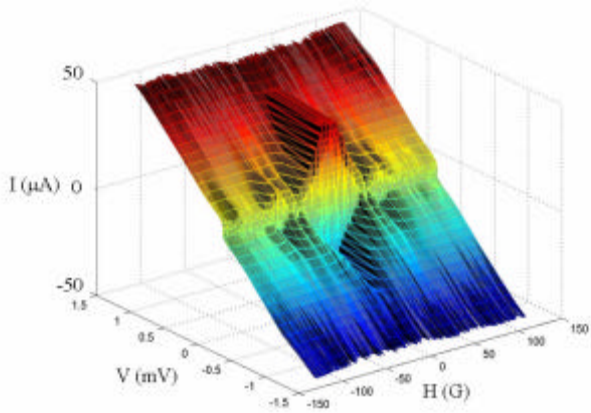

Fig. 2 Current-voltage curves as a function of the magnetic field.

## 3. DISCUSSION

By performing the ensemble average of $G_N$ over $H$, additional insights can be gained. The autocorrelation function is defined as:

$$DG(\delta V) = 1/N_V \sum_V \sqrt{<(G_{V+dV}(H) - \overline{G_N})(G_V(H) - \overline{G_N})>_H} =$$

$$= 1/N_V \sum_V \sqrt{<dG(dV)dG(0)>} \quad (1)$$

$<dG(\delta V) dG(0)>$ is reported in Fig. 4 at $T = 300$ mK. The plot is the result of an average over various series of measurements, to suppress sample specific effects. An energy scale arises naturally, in correspondence to the zero of the function, that we identify with the Thouless energy (~1 meV), in analogy to similar studies of normal metal and semiconductor systems [13,14]. This is proportional to the inverse time an electron spends in moving coherently across the mesoscopic sample. It is quite surprising that quasiparticles seem to travel coherently across the junction even if $V >> E_C$. In Fig. 3a we report also the fit obtained by calculating the contribution to the autocorrelation due to diffusions [15,16] in the limit $k_B T >> h/t_j$.

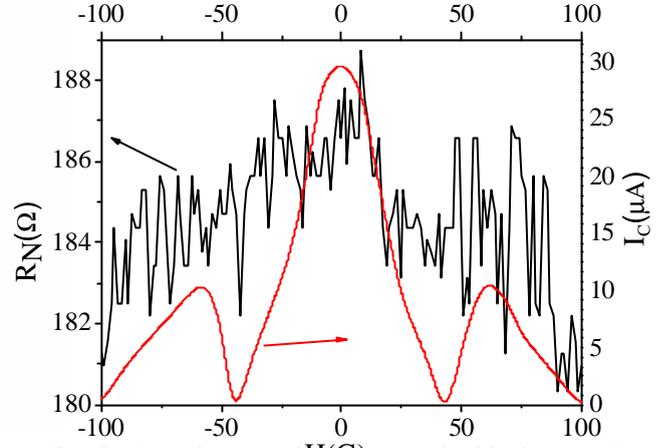

Fig. 3. Normal state resistance $R_N$ and critical current $I_C$ are reported as a function of the magnetic field H for T = 300 mK. $I_C$ (scale on the right) is characterized by Fraunhofer-like dependence. $R_N(H)$ (scale on the left) is characterized by non periodic fluctuations.

We have found that both temperature and bias voltages, up to 30 meV, reduce the amplitude of the CFs and of $e_c$, but they do not wash out the correlations. $\sqrt{<(dG/\overline{G_N})^2>}$ drops with temperature, as probed from the experiment.

Given the thermal diffusion length $L_T \sim 0.13$ μm at 1 K (with $D \sim 20-24$ cm$^2$/s for YBCO [17]), the condition $L_x < L_T$ is certainly met, thus confirming that $e_c$ is the relevant energy scale. The phase coherence length exceeds the size of the microbridge and, therefore, thermal decoherence can be ruled out. Hence, not only the interference of tunnelling currents takes place, which provides the Fraunhofer pattern of Fig. 2, but also the quantum interference of electrons returning back to the junction in their diffusive motion at the GB.

As compared with Al/Al-oxide mesoscopic constrictions, which were studied in the late '90s [13], we have here occasion to monitor HTS quasiparticles driven out of equilibrium by the bias. In our case, the *qp* phase coherence time $t_j$ seems not to be limited by energy relaxation due to voltage induced non-equilibrium [16]. Mesoscopic resistance oscillations are found even for $eV > e_c$, indicating that *qp*s do not loose coherence at low temperatures, while diffusing across the normal conducting bridge.

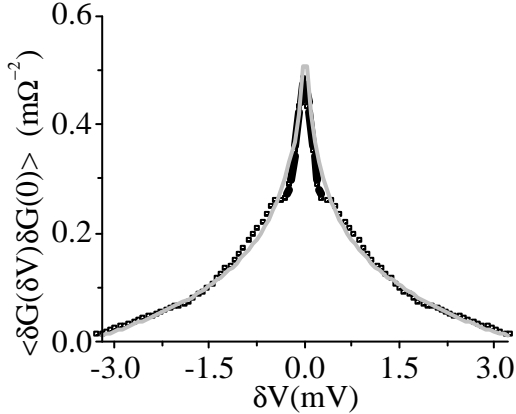

Fig. 4 *<dG(δV) dG(0)>* related to the autocorrelation function of Eq. 1 is derived from experimental data at *T* = 300 mK (black squares). The fit obtained by calculating the contribution to the autocorrelation due to diffusions [16,11] (grey line).

As a final remark, our measurements show that $e_c$ is the relevant energy scale for the supercurrent as well. Indeed, we find that $eI_cR_N$ and $e_c$ are of the same order of magnitude, in agreement with the typical values measured in HTS JJs [2,3], and the results on diffusive phase coherent normal-metal SNS weak links [18,19]. This feature gives additional consistency to the whole picture, relating the critical current, which is mediated by subgap Andreev reflection, with the transport properties at high voltages. The coherent diffusive regime across the SNS channel of our GB junctions persists even at larger voltage bias $D > eV > e_c$ and is the dominating conduction mechanism.

Hence, microscopic features of the weak link appear as less relevant, in favor of mesoscopic, non local properties. The fact that the dominant energy scale is found to be $e_c$ ~ 1 meV sets a lower bound to the relaxation time at low *T* for *qp*s of energy even 30 times larger, of the order of picoseconds.

## 4. CONCLUSIONS

In conclusion, we have given evidence of conductance fluctuations in HTS grain boundary Josephson junctions constricted by one single micro-bridge at low temperatures. Conductance fluctuations are the signature of a coherent diffusive regime. Our results seem to suggest an unexpectedly long *qp* phase coherence time, which represents another strong indication that the role of dissipation in HTS has to be revised [4,5]. Possible applications could exploit the self-protected fingerprints on the conductance on a large range of applied voltages.
We plan to search for similar phenomena in artificially controlled nano-bridges, fabricated through with nano-lithography techniques. Preliminary steps have been successfully carried out. We expect from these studies also major feedbacks for an understanding of transport processes across grain boundaries and barriers in HTS junctions.

## 5. ACKNOWLEDGMENTS

This work has been partially supported by MIUR under the project "Quantum effects in Nano-structures and Superconducting devices".